# On-Line Tests

**University Assistant Florentina Anica Pintea**
„Tibiscus" University of Timişoara, România

ABSTRACT. This paper presents an interactive implementation which makes the link between a human operator and a system of a administration of a relational databases MySQL. This application conceived as a multimedia presentations is illustrative for the way in which the transfer and the remaking of the information between the human operator, the module of data processing and the database which stores the informations can be solved (with help of the PHP language and the web use) .

## 1. Technical information about Online Tests

The application presented by this project is actually an illustration of an interactive programming category able to perform the data taking over from a database and then to process them. In principle the connection to a database (in case of present Mysql) can be done simply by help of PHP modules.

The .html pages were created by the help of Microsoft FrontPage, and those containing php codes were saved by the php extension. All the operations - insertions, modifications, data deleting – operated on the data base are created by PHP. The index of the application is made by more frames.

In order to have a more dynamic and attractive page, we used a few animations created by Adobe Photoshop, and for the colors of the texts we used htlm font's html the CSS files. (Cascading Style Sheets).

There are two types of users that may access this sit. One of them is the administrator who can be a person or some teachers who can access and implement information from and in the data base; the other type is the normal user who can have a test and also can check how many tests he/she has had and what results does he/she obtained for each. No user may have a test without having an account.

Running on the local host server, maintaining the database, as well as its manipulation can be easy done by using Phpmyadmin, running on local host as the root@localhost.

The supervisor can input tests, visualize all the users who have had a tests, can delete certain tests, modify all kind of registrations, therefore he has total control the database.

## 2. Defining of the application.

The application consists in defining the database necessary for the stock of the information about domains, sub domains, questions, and answers from tests and in the creation of a users interface. The most important thing in an application created for a user is its interface with the user. The interface with the user must present in a manner as explicit as possible the functions of the application it interfaces.

Firstly, the PHP, Mysql and Apache software must be installed, and the Mysql and Apache servers must be started. We need a text editor, a good example being Winsyntax or Notepad.

The next important step in the realization of the application is the creation of necessary database for the data stock. Then, with help of the PHP language data from the BD will be taken over and displayed as needed.

## 3. The connecting-up to the database

When working with the information from the database through the PHP code, it should be specified what database is used, meaning we have to connect us to the database of which information we want to use.

As more files that gather information from the database have been used, we used for the connection a script „connection_bd.php" by which the connecting-up to BD can be achieved and which can be included whenever needed.

The lines of the file „connection_bd.php" are:

```php
<?php
$connection = mysql_connect('localhost')
            or die ("Connection error");
$name_bd = mysql_select_db('tests')
            or die ("Database selection Error");
?>
```

where:
- mysql_connect("local host") makes the connecting-up to the Web server, and if it is identified by Id and a password, the statement shall be of the kind: mysql_connect( „local host", „id", „password");
- die is function for error and it displays on-screen the desirable message;
- mysql_select_db ('tests') operates the selection of the „tests" database only if the connection to the Web server was successful.

The inclusion of script „connection_bd.php" in each file, where the connection to the data base is needed, will be done by one of the commands:
- include("name_file")
- require("name_file")

In the present work the command:
Include ('conectare_bd.php');
was used.

The files which will be included can contain everything that a PHP script, HTLM labels, PHP functions, PHP class could normally contain. If these instructions appear more frequent in a script, the file will be always included.

## 4. The projection of the database

When creating the database, a process of identification and of organization of the columns, the grouping of the columns in tables (entities) and the settlement of the relations among these must be done.

The database for the interactive application „Online Tests" is a relational database consisting of ten tables. The connections between tables are realized by means of the correspondence between the primary keys of the tables and the external keys of other tables.

This database is created using Phpmyadmin:

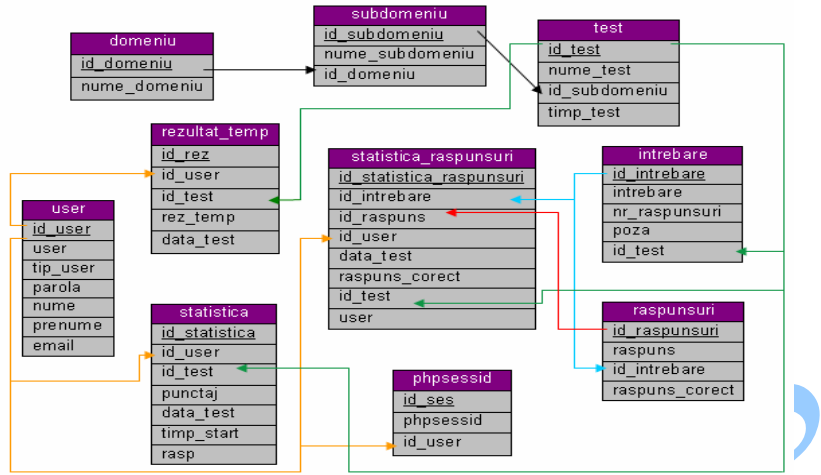

Figure 1. The relation between the tables of the database „test"

## 5. The operation at the level Supervisor

We need a section of administration by which to add or modify the access options for the users of the database, as well as for the access to the addition-modification-deleting of data regarding the tests or the users (can erase the tests already had by a certain users).

In the moment of connection, the supervisor will have access a menu, from which he could make the option he/she wants.

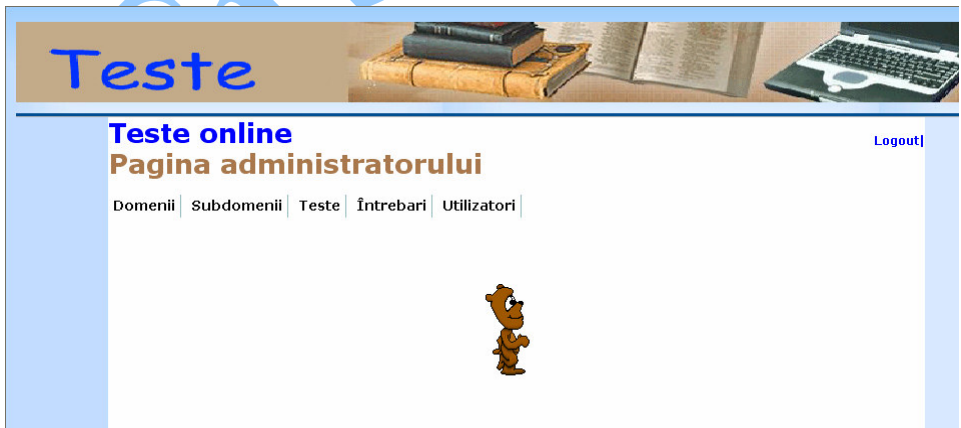

Figure 2. The interface with the Supervisor – options

## 6. The operation at the User level

The user must firstly create an account and only then he/she can have a test or the needed tests. If he has already had an account and he has forgot the password he/she can get it back...

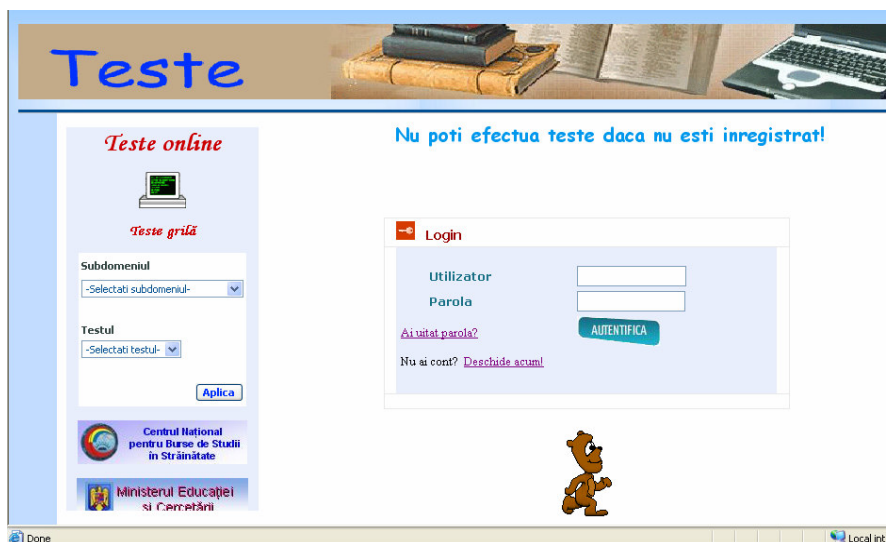

Figure 3. The page of User

After the user has registered, he can have the tests and see his/her last 20 done tests. There is a warning message which draws the user's attention on the fact that he/she cannot turn back to a question if he/she passed to another one.

From left side, the domains and the afferent sub-domains can be chosen as well as correspondent tests. On the page with the test, the questions and answers are displayed, and the user has to choose from the selection boxes the correct answers. On that page data on the user are also displayed – the user, name, and first name – as well as different data such as: the present date, the time for starting the test, the final time. Each the test has a time limits. If it is exceeded the test is interrupted at the current question and the score is done.

If the given answers are correct, the temporary score is count up and displayed on-screen.

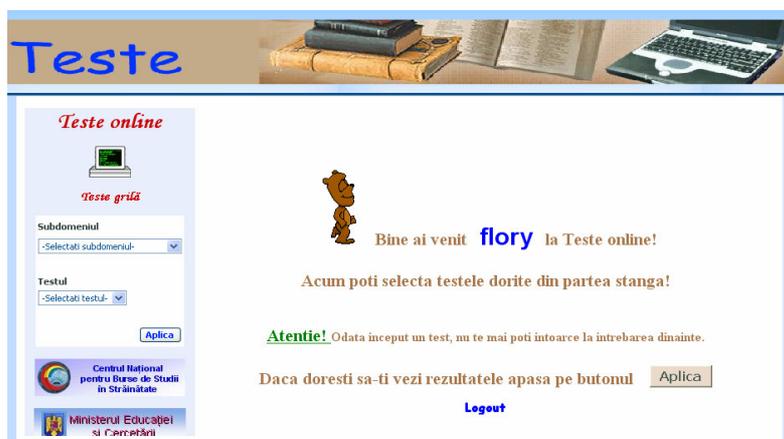

Figure 4. User connected

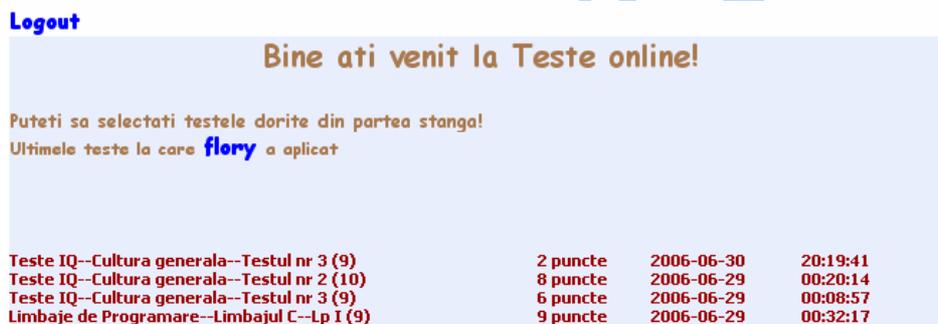

Figure 5. The done test

On the left part, we have and some links which point to the web pages about education.

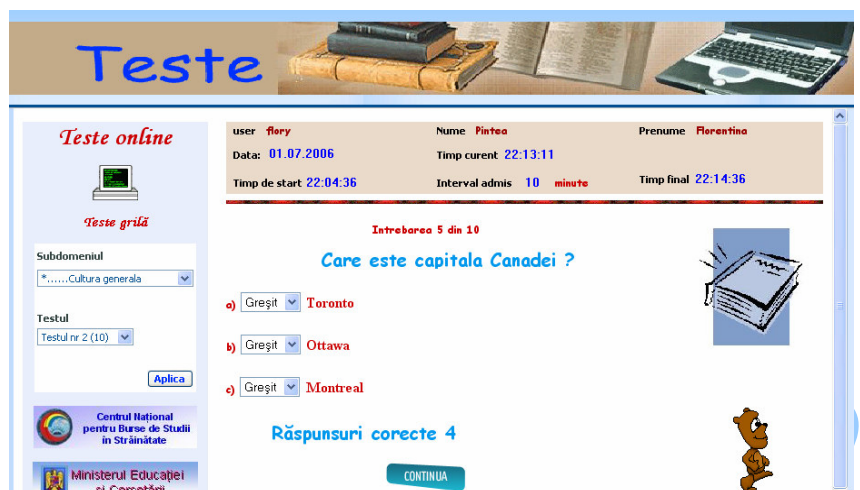

Figure 6. Display the current query

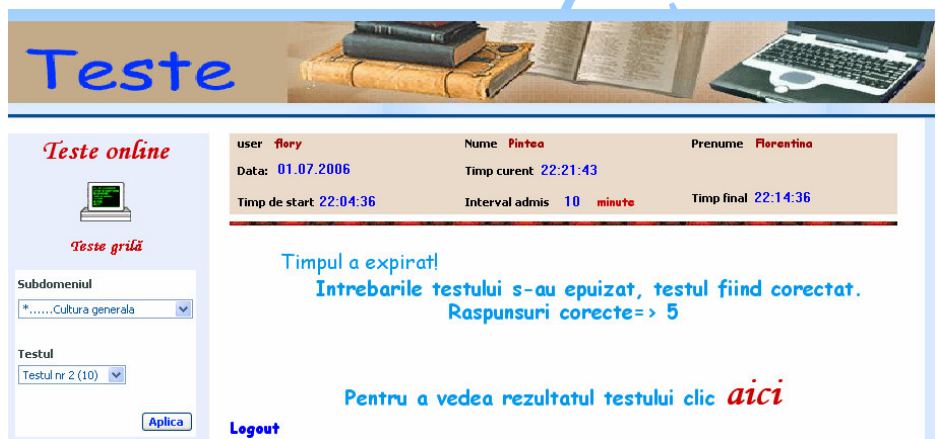

Figure 7. Time expired

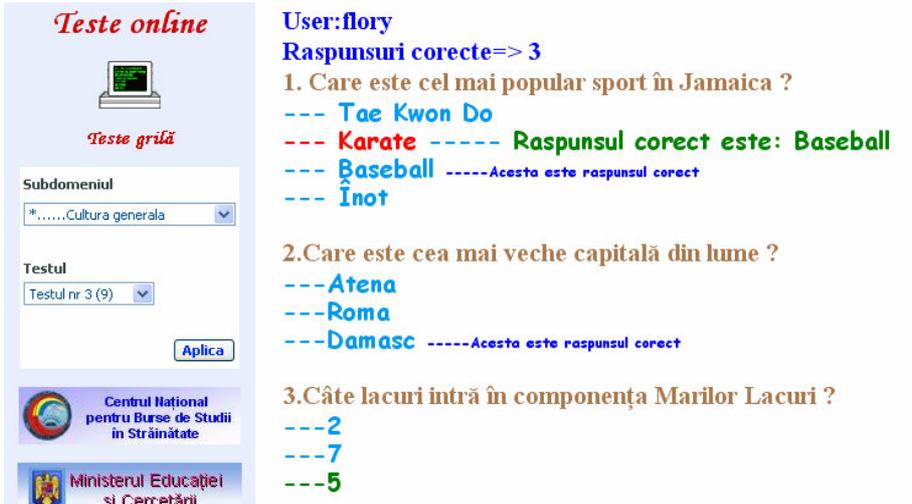

Figure 8. Displayed result of the test